\documentclass[prb,floatfix,twocolumn,preprintnumbers,showpacs,amsmath,amssymb]{revtex4}
\usepackage{graphicx}
\usepackage{dcolumn}
\usepackage{bm}

\begin{document}

\title{A microscopic theory for the incommensurate transition in TiOCl}
\author{Diego Mastrogiuseppe}
\affiliation{Facultad de Ciencias Exactas Ingenier{\'\i}a y
Agrimensura, Universidad Nacional de Rosario and Instituto de
F\'{\i}sica Rosario, Bv. 27 de Febrero 210 bis, 2000 Rosario,
Argentina.}
\author{Ariel Dobry}
\affiliation{Facultad de Ciencias Exactas Ingenier{\'\i}a y
Agrimensura, Universidad Nacional de Rosario and Instituto de
F\'{\i}sica Rosario, Bv. 27 de Febrero 210 bis, 2000 Rosario,
Argentina.}
\date{\today}

\begin{abstract}
We propose a microscopic mechanism for the incommensurate phase in
TiOX compounds. The model includes the antiferromagnetic chains of
Ti ions immersed in the phonon bath of the bilayer structure.
Making use of the Cross-Fisher theory, we show that the
geometrically frustrated character of the lattice is
responsible for the structural instability which leads the chains
to an incommensurate phase without an applied magnetic field. In
the case of TiOCl, we show that our model is consistent with the
measured phonon frequencies at $T=300K$ and the value of
the incommensuration vector at the transition temperature.
Moreover, we find that the dynamical structure factor shows a
progressive softening of an incommensurate phonon near the zone
boundary as the temperature decreases. This softening is accompanied by a broadening of
the peak which gets asymmetrical as well when going towards the
transition temperature. These features are in agreement with the
experimental inelastic X-ray measurements.
\end{abstract}
\maketitle

\section{Introduction}
In 1955, when writing an introductory solid-state textbook, \cite{Peierls}
Rudolph Peierls discovered that no one-dimensional metal could
exist due to the electron-phonon coupling. Indeed, a half-filled
metal is unstable towards a lattice dimerization. The system
undergoes a metal-insulator transition because the loss of
elastic energy is made up for by the electronic energy acquired when
the gap is opened. A quite similar process takes place in a
one-dimensional antiferromagnet, giving rise to the so-called
'spin-Peierls' (SP) transition. \cite{reviewSP}
The lattice dimerizes and the antiferromagnetic quasi long range order is replaced by a gapped singlet state.
From this point of view, no one-dimensional antiferromagnetic system should be stable.

In spite of this  general prediction, only few spin-Peierls
materials  were found. Most of them were observed in the early 70s
in some organic charge transfer systems. However, direct spectral
characterization was lacking due to the unavailability of large
crystals for inelastic neutron scattering measurements. In this
context the discovery of CuGeO$_3$ in 1993, the first inorganic
spin-Peierls system, renewed the interest in this subject. Large
enough crystalline samples of CuGeO$_3$ could be obtained to
undertake a detailed spectral characterization. An important
conclusion of these studies was that in CuGeO$_3$, the
nonadiabatic character of the phonons \cite{Uhrig} is so important that the SP
transition is not driven by a softening of a precursive phonon
mode. It has been shown that an extension of the canonical theory
of Cross-Fisher \cite{CF} could explain this feature. \cite{GW}
Moreover,  the dispersion relation of the phonons in the direction
perpendicular to the magnetic chains should be taken into account
in a nonadiabatic SP system like CuGeO$_3$. \cite{DCR}

The recent discovery of a spin-Peierls transition in the TiOX (X = Cl, Br)
family has opened new questions about this spin-Peierls paradigm.
The essential building blocks of these compounds are bilayer structures of
magnetically active Ti ions connected by O ones.
The position of a Ti ion in a layer is shifted with
respect to the other in the neighboring layer, forming something
like an anisotropic triangular structure. In fact TiOCl was initially thought to be
a candidate for a \textit{resonating valence bond} state, \cite{beynon} but it
has been found to be mainly a one-dimensional
magnetic system \cite{shaz} with the Ti d$_{xy}$ orbitals pointing toward each other in the
crystallographic $b$ direction. \cite{seidel}
The high temperature magnetic susceptibility is well described by the
Bonner-Fisher curve, indicating a nearest neighbor magnetic exchange $J\sim 660K$. \cite{seidel}

The phase diagram of TiOCl does not correspond to a canonical SP
system in the sense that an incommensurate intermediate phase
appears between the high temperature uniform phase and the low T
dimerized phase. \cite{seidel,imai} The transition temperatures
are found to be $T_{c1}\sim 66K$ and $T_{c2}\sim 92K$. In Ref.
\onlinecite{imai} it is also reported a very large energy gap of about
$430K$ in the low temperature phase and a pseudo spin gap below
$135K$. The order of both transitions is still under debate.
\cite{ruckamp,schonleber2,fausti}

The origin of the intermediate phase is controversial and not yet
well understood. As the position of a Ti ion in a chain is shifted
with respect to another one in the neighboring chain, it is
plausible to speculate that some type of competition between the
in-chain and out-of-chain interactions could be the origin of the
incommensurate phase. This was the idea of R\"uckamp \textit{et al}.
\cite{ruckamp} who proposed a Landau theory for the incommensurate
transition. This phenomenological theory includes the tendency to
dimerize of each chain and the coupling of the order parameter
with the neighboring chain. As a result the incommensurate phase
is accounted. However the sign and the value of the parameters are
phenomenologically chosen and the connection with the underlying
microscopic theory is not clear.

In view of the previous discussion, in this paper we propose a
simplified microscopic model which accounts for the transition from
the uniform to the incommensurate phase in TiOCl. Our model
contains the relevant magnetic interaction, the phonons and the
spin-phonon coupling. The paper is organized as follows:
In Sec. \ref{model} we present the model and show that the essential ingredient 
leading to the incommensurate transition is that the phononic 
dispersion of the modes near $\pi$ is linear. In Sec. \ref{tiocl} we apply the model to TiOCl.
We start by fitting the parameters in order to account for the measured phononic frequencies
and the value of the incommensurate wave vector at the transition temperature.
After that, we obtain the dynamical structure
factor of the phonons and compare it with the one obtained in X-ray experiments. The softening 
of a phonon and the loss of spectral weight when the transition approaches is correctly accounted.
Finally, in Sec. \ref{discussion} we discuss our results in comparison with another approach and summarize them.

\section{\label{model} Model and incommensurate transition}
\begin{figure}[ht]
\includegraphics[width=5cm]{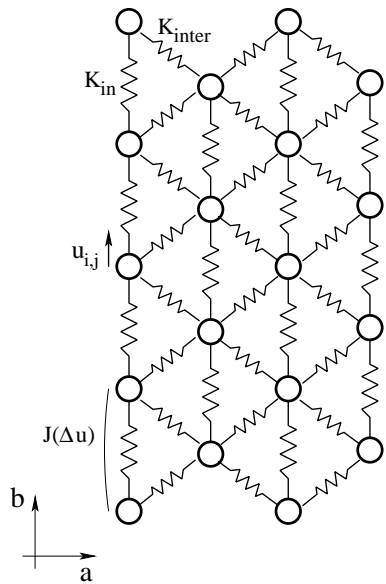}
\caption{\label{modelph2d} Schematic representation of our
simplified model. Only Ti atoms are included over the $ab$ plane.
$K_{in}$ and $K_{inter}$ are the harmonic force constants acting
when two neighbor atoms in the same chain and in neighboring chains respectively,
move from their equilibrium position.}
\end{figure}

 We include the Ti atoms on the bilayer
structure which interact by harmonic forces as depicted in Fig. \ref{modelph2d}.
For simplicity we consider ionic displacements only in the direction of the magnetic chains.
As the measured
magnetic susceptibility of TiOCl is well reproduced by a 1D
Heisenberg model, we assume that the Ti atoms are magnetically
coupled only along the $b$ direction. Moreover, as a direct
exchange seems to be the dominant Ti-Ti interaction, we assume
that $J(\Delta u)$ is modulated by the movement of nearest neighbor Ti ions in
the chain direction. Our spin-phonon Hamiltonian reads:

\begin{eqnarray}
\label{hspinph}
H&=&H_{ph}+H_{s}+H_{sph},\\
H_{ph}&=&\sum_{i,j}{\frac{P_{i,j}^2}{2m}} +  \sum_{i,j}\lbrace\frac{K_{in}}{2}(u_{i,j}-u_{i+1,j})^2\nonumber\\
&+&\frac{K_{inter}}{2}\left[( u_{i,j}-u_{i,j+1})^2 +( u_{i,j}-u_{i+1,j-1})^2\right]\rbrace,\nonumber\\
H_s&=&J \sum_{i,j}  \textbf{S}_{i,j} \cdot \textbf{S}_{i+1,j},\nonumber\\
H_{sph}&=&\sum_{i,j} \alpha (u_{i+1,j}-u_{i,j})\,
\textbf{S}_{i,j} \cdot \textbf{S}_{i+1,j},\nonumber
\end{eqnarray}

\noindent where $P_{i,j}$ is the momentum of the atom $i$ of the chain $j$, $u_{i,j}$ are the displacements from the equilibrium positions along the direction $b$ of the magnetic chains, $\textbf{S}_{i,j}$ are spin-$\frac12$ operators with exchange constant $J=J(\Delta u=0)$ along the b-axis of a non-deformed underlying lattice, $\alpha = (\textrm d J(\Delta u)/\textrm d \Delta u)|_{\Delta u = 0}$, and $K_{in}$ and $K_{inter}$ are the in-chain and inter-chain harmonic force constants as shown in the Fig. \ref{modelph2d}.

\noindent Representing the atom displacements and momenta through phonon normal coordinates, we can write the phonon dependent parts of the Hamiltonian as:

\begin{eqnarray*}
 H_{ph}&=& \sum_{\textbf{q}}{\hbar \Omega(\textbf{q})\left(a^\dagger_{\textbf{q}}\, a_{\textbf{q}} + \frac{1}{2}\right)} , \\
 H_{sph}&=& \frac{1}{\sqrt{N}} \sum_{\textbf{q}}{g\,(1-e^{{\rm i} b q_y}) } Q(\textbf{q}) \sum_{i,j}{e^{{\rm i}\textbf{q}\cdot \textbf{R}_{i,j}} \textbf{S}_{i,j}\cdot \textbf{S}_{i+1,j}},
\end{eqnarray*}

\noindent where $\Omega(\textbf{q})$ is the phonon dispersion relation which reads
\begin{eqnarray}
\label{Wq}
\Omega^2(\textbf{q}) = \frac{4}{M} [K_{in} \sin ^2{\frac{q_y}{2}} + K_{inter} (1-\cos{\frac{q_x}{2}} \cos{\frac{q_y}{2}})],
\end{eqnarray}
\noindent for our model and
$g=\frac{\alpha}{\sqrt{M}}$. The quantized phonon normal coordinates are defined by
\begin{eqnarray*}
Q(\textbf{q}) = \sqrt{\frac{\hbar}{2 \Omega(\textbf{q})}} (a^{\dagger}_{-\textbf{q}} + a_{\textbf{q}}).
\end{eqnarray*}

\noindent It will be also useful to introduce a dimensionless spin-phonon coupling constant defined as in Ref. \onlinecite{CF}:
\begin{eqnarray*}
\lambda=\frac{4g^2}{\pi J \Omega^2(q_y=\pi)}.
\end{eqnarray*}

To compare with X-ray scattering data, the dynamical structure factor  $S(\textbf{q},\omega)$ has to be determined. It is related to the phononic retarded Green's function \cite{mahan} by
\begin{eqnarray*}
S(\textbf{q},\omega)=-\frac{\Omega(\textbf{q})}{\pi}\frac{\mathrm{Im}\;\mathcal{D}^{ret}(\textbf{q},\omega)}{1-e^{\beta
\omega}}, 
\end{eqnarray*}

\noindent where
\begin{eqnarray}
\mathcal{D}^{ret}(\textbf{q},\omega) = \frac{\mathcal{D}^0(\textbf{q},\omega + {\rm i}\delta)}{1-\mathcal{D}^0(\textbf{q},\omega + {\rm i}\delta) \Pi(\textbf{q},\omega + {\rm i}\delta)}. \label{Gret}
\end{eqnarray}

\noindent $\mathcal{D}^0$ denotes the non-interacting phonon Green's function defined by
\begin{eqnarray*}
\mathcal{D}^0(\textbf{q},{\rm i}\omega_n) = \frac{2}{({\rm i}\omega_n)^2 - \Omega ^2(\textbf{q})},
\end{eqnarray*}

\noindent and $\Pi(\textbf{q},\omega)=\Pi(q_y,\omega)$ is the phonon self-energy.
In the following, we treat the spin-phonon interaction using a random phase approximation (RPA) for the phonon self-energy term along with the expression obtained by Cross and Fisher \cite{CF} using bosonization for the dimer-dimer correlation function in the Heisenberg model:
\begin{eqnarray*}
\Pi(q_y,\omega) &=& -\frac{0.37}{T}|(1-e^{{\rm i} q_y b})g|^2 I_1\left(\frac{\omega+\frac{\pi}{2}Jb(q_y - \pi) }{2\pi T}\right)\\
&\times& I_1\left(\frac{\omega-\frac{\pi}{2}Jb(q_y - \pi) }{2\pi T}\right),
\end{eqnarray*}

\noindent with
\begin{eqnarray*}
 I_1(k) = \frac{1}{\sqrt{8\pi}} \frac{\Gamma\left(\frac{1}{4} + \frac{1}{2} {\rm i}k \right)}{\Gamma\left(\frac{3}{4} + \frac{1}{2} {\rm i}k \right)}.
\end{eqnarray*}
Let us start by identifying the structural transition of the model
by searching the poles of the retarded Green's function when
$\omega=0$ over the ($q_x,q_y$) plane of the first Brillouin zone.
They are given by the equation of the zeros in the denominator of
Eq. (\ref{Gret}) which reads:
\begin{eqnarray}
\label{Tq}
 \Omega^2(\textbf{q})+2 \Pi(q_y,0)=0.
\end{eqnarray}
From this equation we obtain $T(\textbf{q})$, the temperature where
the renormalized frequency vanishes for each $\textbf{q}$. The highest
of these temperatures signals the \textit{\textbf{q}} wave vector of
the structure that will be developed at that T. For an isolated chain this
transition takes place at $\textbf{q}=(0,\pi)$, which is the usual
spin-Peierls transition towards the dimerized phase.

How does this situation change when
the system acquires a transversal dispersion due to an elastic coupling between the chains? Certainly, if the chains
arrange in a rectangular geometry (atoms in different chains align in phase),
the dispersion of the phonons in the chain direction has a local maximum at 
$q_y=\pi$ and the $T(\textbf{q})$ given by (\ref{Tq}) has a maximum at $\textbf{q}=(0,\pi)$.
Thus, no complete softening of an  incommensurate mode is found.

A crucial consideration here is that the dispersion given by Eq. (\ref{Wq}), when expanded near
$q_y=\pi$, reads:
\begin{eqnarray*}
\label{Wqapp}
\Omega^2(\textbf{q}) &\sim& \frac{4}{M} [K_{in}  + K_{inter} (1-\frac{\delta}{2}\cos{\frac{q_x}{2}} )],\\
\delta&\equiv&\pi-q_y, \nonumber
\end{eqnarray*}
i.e. it increases linearly near $\pi$. Plugging this expression in Eq. (\ref{Tq}) and expanding up to order $\delta$ we obtain:
\begin{eqnarray}
\label{Tlinear}
T(\delta)= \frac{g^2 M  |a_0|}{K_{inter}+K_{in}} \left[1+\frac12 \frac{ K_{inter}  \cos(\frac{q_x}{2}) \delta}{K_{inter}+K_{in}}\right],
\end{eqnarray}
where it was enough to calculate $\Pi(q_y,0)$  at $q_y=\pi$ as $\Pi(\pi,0)=\frac{2 a_0 g^2}{T}$, with $a_0=-0.257743$,
because the first correction is of second order in $\delta$.

The temperature given in (\ref{Tlinear}) decreases with $q_x$ and increases for increasing $\delta$. Therefore an instability is predicted
at $q_x=0$ and $q_y$ shifted from $\pi$. The next orders of $T(\delta)$ stabilize a local maximum 
at a given $\delta$. Thus, due to the linear dispersion
of the relevant phonon mode, the model (\ref{hspinph}) has a transition from a uniform to an incommensurate phase by lowering the temperature.
The physical reason why this transition appears in the triangular arrangement and not in the rectangular one is that the elastic energy that the system 
gains by distorting the lattice in a modulated pattern decreases faster in the first case, when the wave vector of the modulation moves away from $\pi$. This reduction
overcomes the cost in the magnetic free energy the system should pay to separate from the dimerized state. This reduction does not take place in the rectangular case and 
the system never goes to an incommensurate phase but directly to a dimerized one.

\section{\label{tiocl} Application to TiOCl}

Having shown that the uniform-incommensurate transition is present in our model, we will try to fit the parameters to account for the available experimental data of TiOCl.
We intend to reproduce the experimental incommensurate wave vector $q_y \simeq 3.04$ \cite{krimmel,abel} and the renormalized frequency of the phonons near the zone boundary at $300K$ obtained in Ref. \onlinecite{abel} by inelastic X-ray scattering.
In order to do that we have to fix the free parameters of our model, i.e. the elastic constants and the spin-phonon coupling.
Usually the critical temperature is used to fix the spin-phonon coupling. This value could be obtained by numerical solution of Eq. \ref{Tq} once the other parameters are known. The values of $K_{in}$ and $K_{inter}$ are fixed in such a way to fit the phononic frequencies measured by inelastic X-ray scattering at $T=300K$ with the ones obtained from the positions of the peaks of $S(q,\omega)$.

\begin{figure}[ht]
\includegraphics[width=7cm]{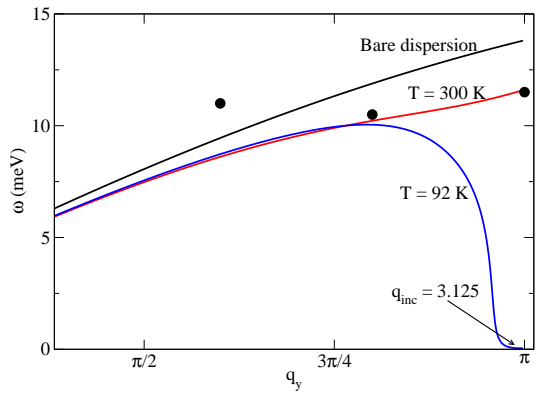}
\caption{\label{wvsqexperim2}(Color online) Evolution with the temperature of the dressed frequency of the phonons along the ($0,q_y$) path. The model parameters are $\sqrt{4K_{in}/M}=6.6\, meV$, $K_{inter}\sim 3.38 K_{in}$ and $\lambda=0.17$. The black dots are the experimental points obtained by inelastic X-ray scattering at $T=300K$. \cite{abel}}
\end{figure}

The value of the in-chain elastic constant was set to $\Omega_{in}=\sqrt{4K_{in}/M}=6.6\, meV$ and the inter-chain one was fixed to $K_{inter}/K_{in}=3.38$. This enables us to fit the experimental frequencies near $q_y=\pi$ at $300K$ with an spin-phonon coupling constant $\lambda=0.17$. The transition temperature is at 92K as in TiOCl. However, using these parameters we obtain an incommensurated wave vector $q_{inc}=3.125$, i.e. it is close to $\pi$ and far away from the experimental value. Our results together with the experimental ones are shown in Fig. \ref{wvsqexperim2}. In this figure we see that, as we move away from the zone boundary, the dispersion curve quickly decreases and separates from the experimental points. Moreover, the bare dispersion curve is very close to the $300K$ one.

\begin{figure}[ht]
\includegraphics[width=7cm]{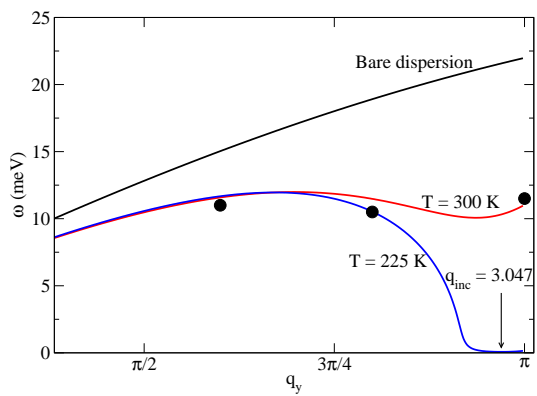}
\caption{\label{wvsqexperim1}(Color online) Evolution with the temperature of the dressed frequency of the phonons along the ($0,q_y$) path. The curves were obtained from the positions of the peaks of the dynamical structure factor. The fitted parameters are $\sqrt{4K_{in}/M}=10.5\, meV$, $K_{inter}\sim 3.38 K_{in}$ and $\lambda=0.41$.}
\end{figure}

We have obtained a quite small value of the incommensuration with those parameters. Therefore, in order to reproduce the experimental one, we increase the spin-phonon coupling $\lambda$. The new value turns out to be $\lambda=0.41$. This change has the undesired effect of increasing the transition temperature up to $T_{inc} \sim 225K$, more than twice the experimental transition temperature. Moreover, if we want to keep adjusting the experimental X-ray phonons at $300K$, we have to increase the value of the bare phonon frequency because of the larger spin-phonon coupling.
The chosen values are $\sqrt{4K_{in}/M}=10.5\, meV$, $K_{inter}\sim 3.38 K_{in}$, i.e. we have kept the ratio $K_{inter}/K_{in}$ unchanged. The results are shown in Fig. \ref{wvsqexperim1}. 
Now we see that the calculated $300K$ curve does not fall rapidly as before and that the experimental point which is the most separated from the zone boundary is also well adjusted. Nevertheless, we have noted that with this set of parameters the transition temperature is overestimated. This problem is also present in the calculations of Abel \textit{et al.}\cite{abel} through an underestimation of the magnetic coupling J. We think that this 
disagreement is due to the adiabatic treatment of the phonons and will be solved by an extension of the results of Ref. \onlinecite{DCR} for the geometry of TiOCl. In this paper it was shown that a theory going beyond the mean field RPA treatment of the phonons by including nonadiabatic effects predicts a reduction of the critical temperature for a given $\lambda$. We plan to undertake this generalization in a forthcoming paper, but let us make a rough estimation of the renormalization of the transition temperature using formula (35) of Ref. \onlinecite{DCR}. Note that the parameter $\Omega_{inter}$ in Ref. \onlinecite{DCR} is not exactly the same as in the present paper. It is in this sense that the following results should be taken as an estimation.  Proceeding in this way we obtain:
\begin{eqnarray*}
 T=T_{ad} \left(1-\frac{1}{\sqrt{1+(\Omega_{inter}/\Omega_{in})^2}}\right)=117K,
\end{eqnarray*}
where $T_{ad}=225K$ is the transition temperature obtained by the adiabatic approach. The non-adiabatic temperature is now very close to the experimental one. Furthermore, let us estimate the energy gap by combining Eqs. (26) and (32) of the same reference:
\begin{eqnarray*}
 \Delta=3.04\lambda J\left(1-\frac{1}{\sqrt{1+(\Omega_{inter}/\Omega_{in})^2}}\right)=427.76K,
\end{eqnarray*}
i.e. we obtain an excellent agreement with the experimental gap of 430K. \cite{imai} From these estimations it can be inferred that the non-adiabatic treatment should solve the problem.

\begin{figure}[ht]
\includegraphics[width=4cm]{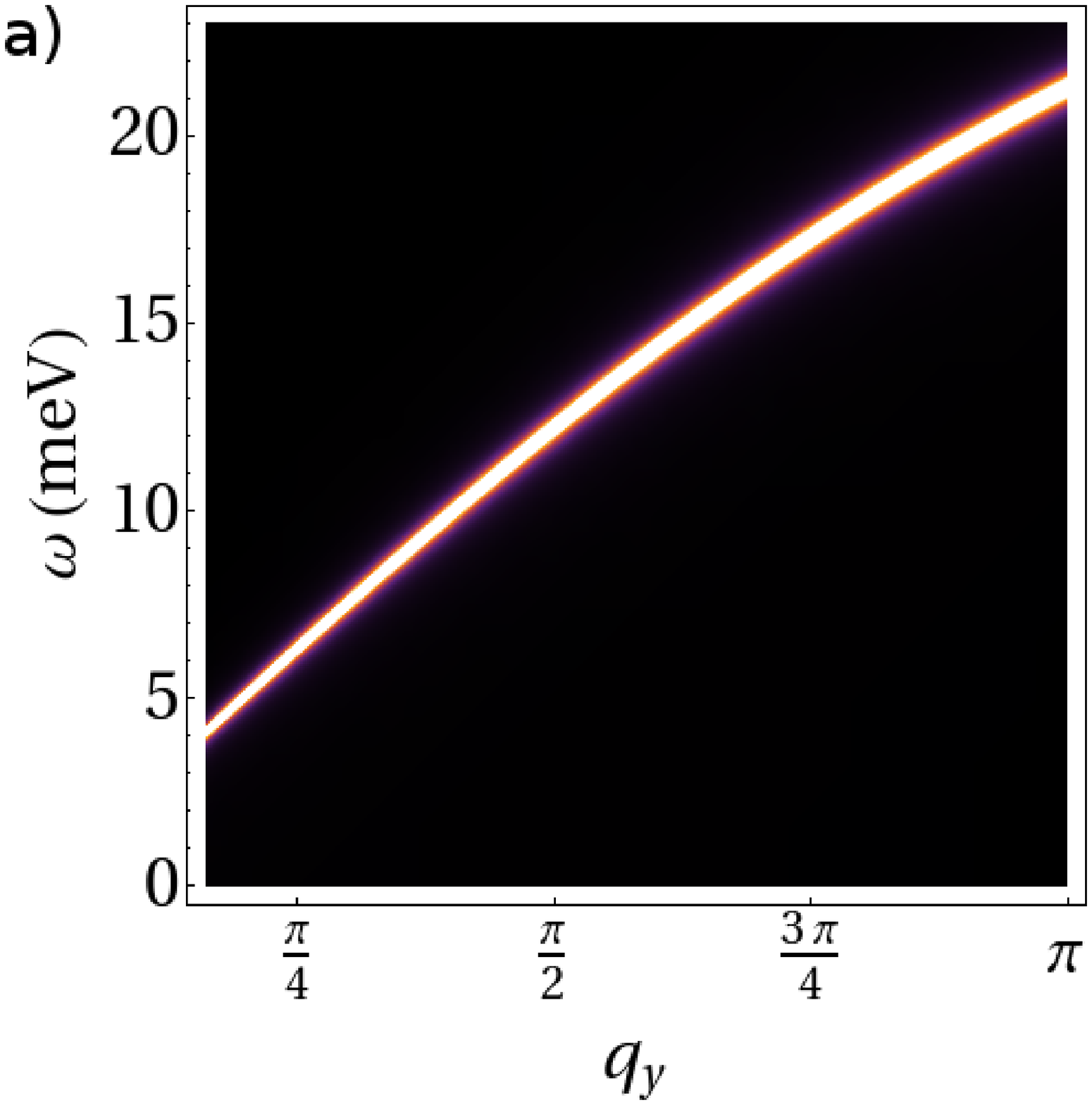}
\includegraphics[width=4cm]{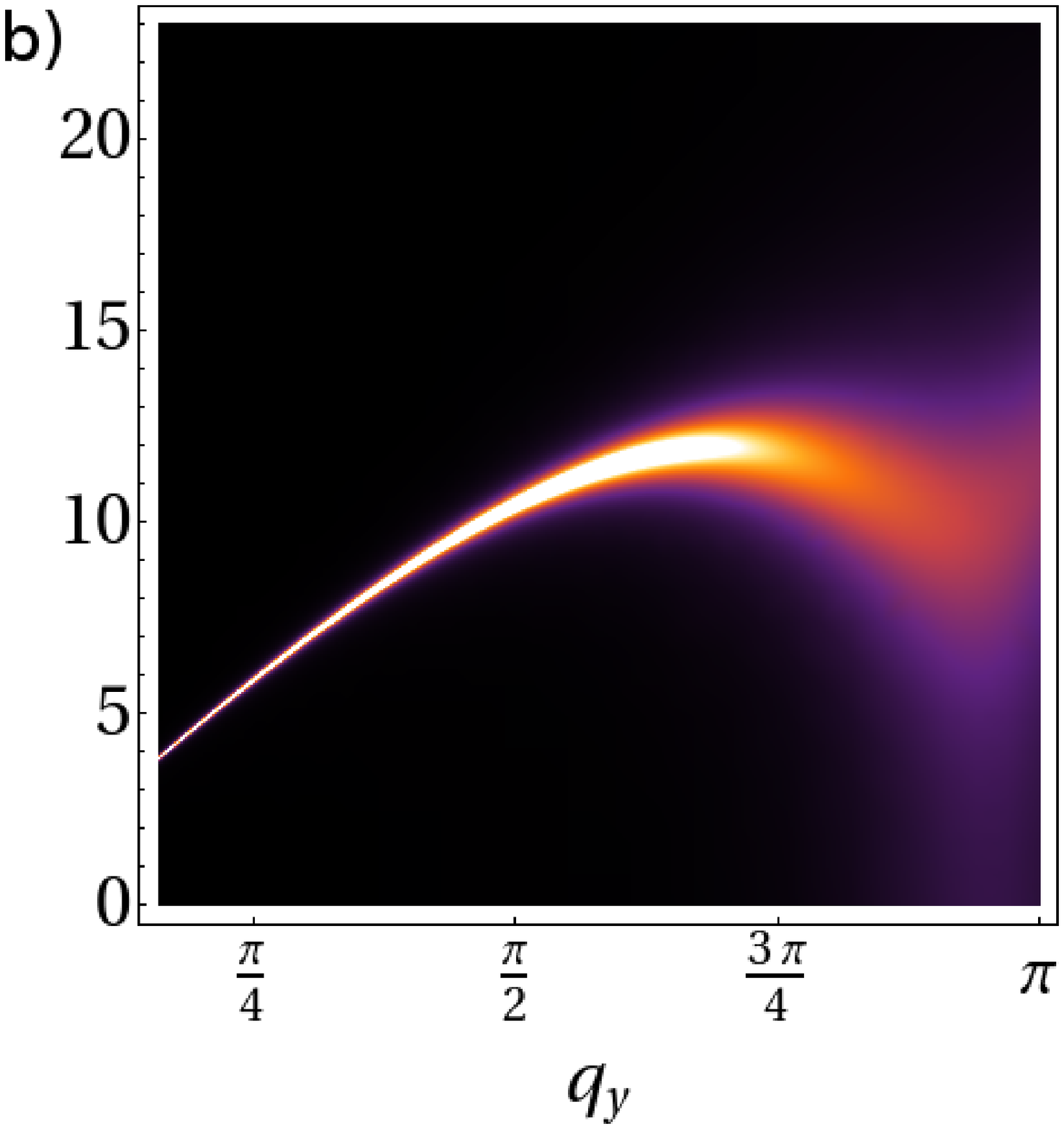}
\includegraphics[width=4cm]{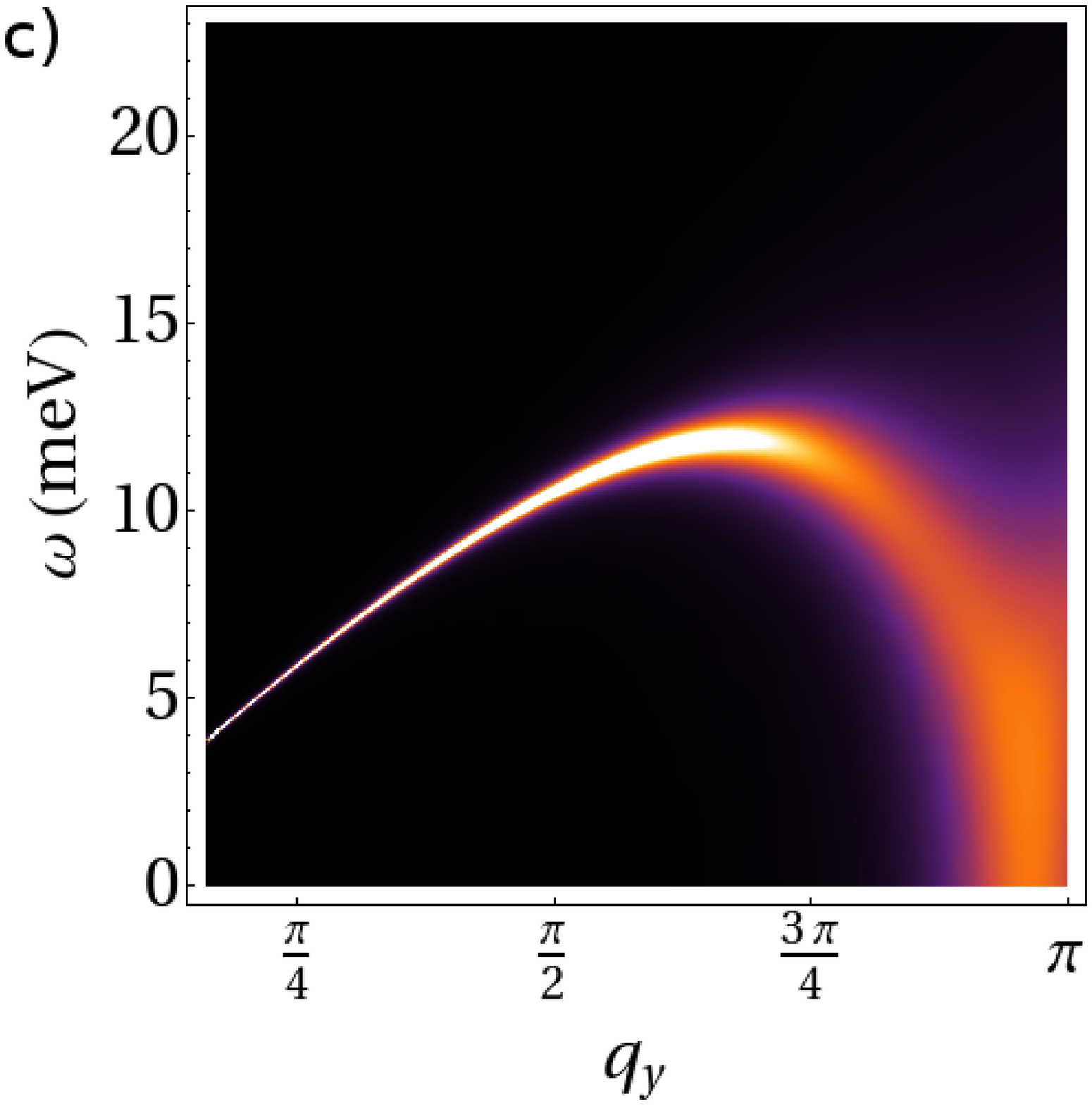}
\includegraphics[width=4cm]{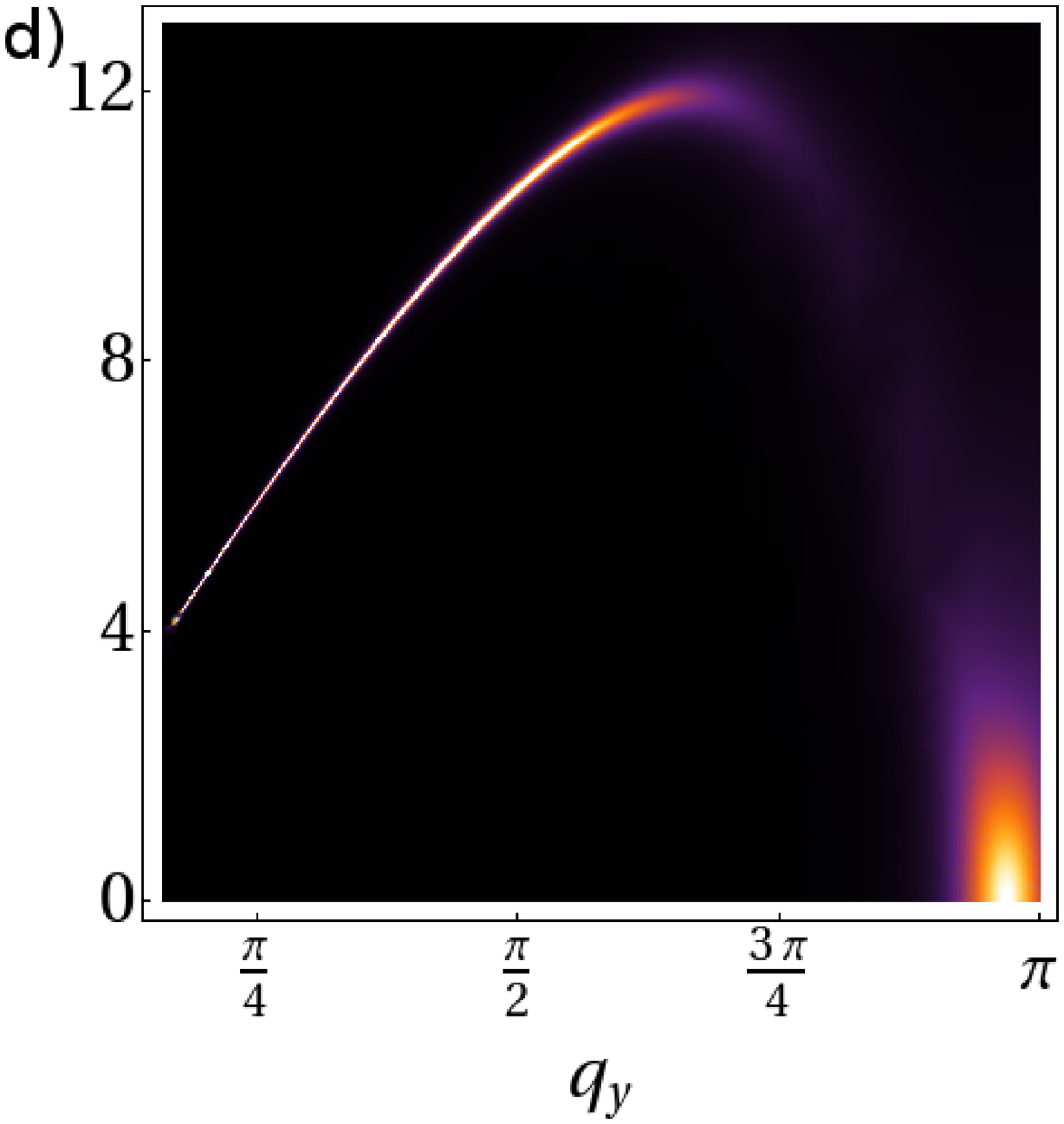}
\includegraphics[width=4cm]{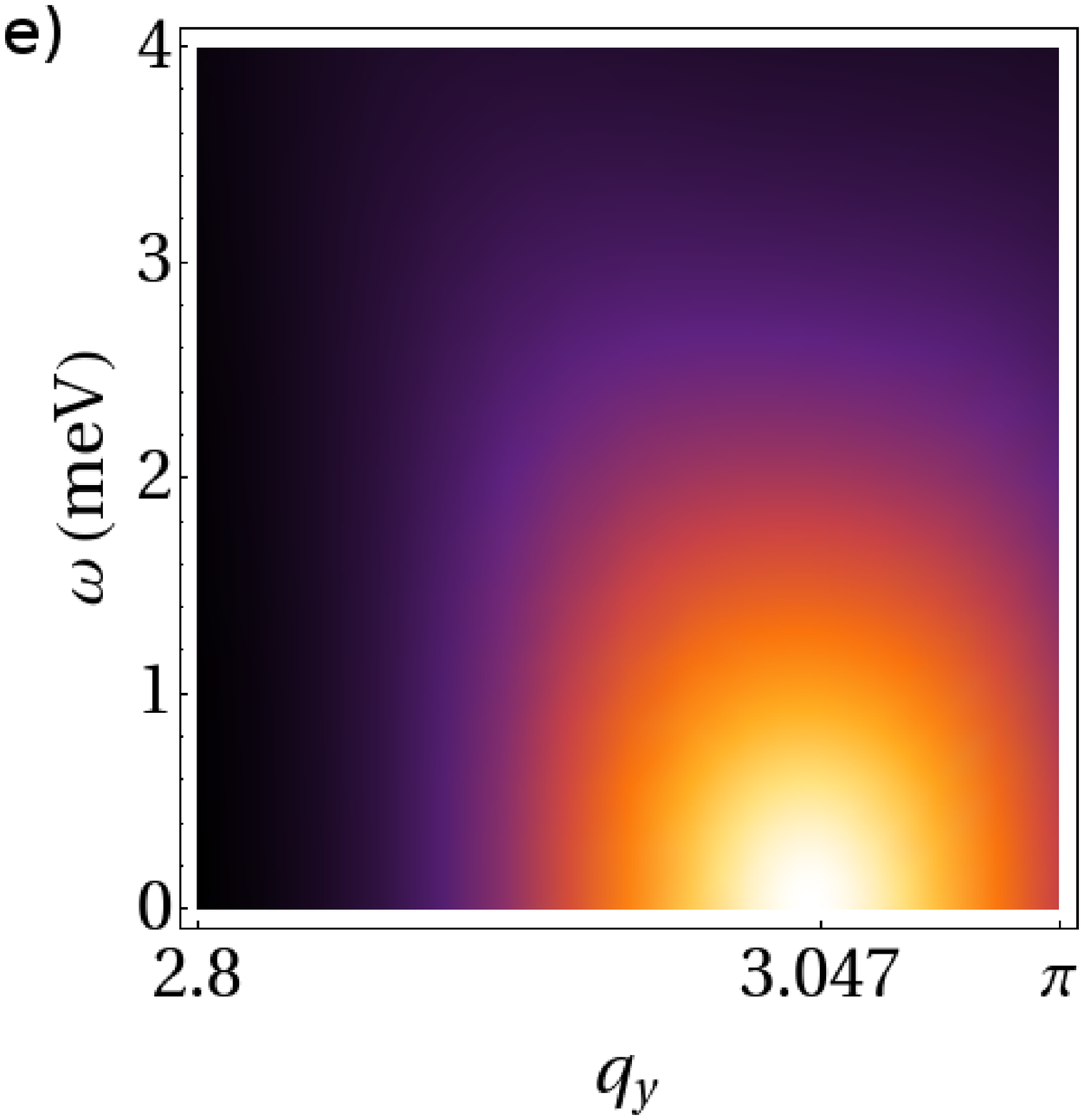}
\caption{\label{s1}(Color online) Contour curves of the  dynamical structure factor as a function of $q_y$ ($q_x=0$) and $\omega$ for different temperatures. (a) $1000K$, (b) $300K$ and (c) $255K$ (d) $225K$ (transition temperature). In (e) we show a zoom of (d) around the transition point. Brighter colors correspond to higher intensities. We see that there is a softening of a group of modes near the zone boundary. In this region, there is a broadening of the peak width with decreasing temperature due to a redistribution of the spectral weight.}
\end{figure}

For the present purpose, let us use the second set of parameters. In Fig. \ref{s1} we show a plot of the dynamical structure factor along the $(0,q_y)$ path of the first Brillouin zone for a range of frequencies and different temperatures.

\begin{figure}[htb]
 \includegraphics[width=7cm]{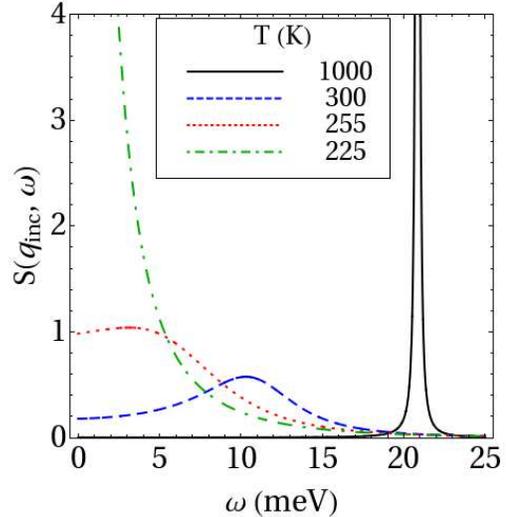}
\caption{\label{s2}(Color online) Dynamical structure factor at $q_{inc}$ for different temperatures. At high T, we see a very definite peak which broadens and gets asymmetrical as temperature decreases. At $T=225K$ (the transition temperature), we observe the divergence of the peak for $\omega=0$.}
\end{figure}

When the temperature is high [Fig. \ref{s1}(a)], we obtain peaks which reproduce the bare
dispersion relation of the model. With decreasing temperatures
[Fig. \ref{s1}(b) to (d)], we observe the trace of the phonon softening
along with a broadening and a height reduction of the peak. This
is due to a redistribution of the spectral weight which is
concentrated again in a defined peak at the transition temperature
for $\omega=0$ and $q_y=3.047$ [Fig. \ref{s1}(d) and zoom in (e)].
This  broadening of the peak indicates that phonons are overdamped
by the interaction with the magnetic excitations. This behavior
has been indeed observed experimentally in Ref. \onlinecite{abel} [see Fig.
9(b) in this reference] and considered there as a disagreement with the
Cross-Fisher theory. Instead, our results show that due to the interaction of the bare phonons
with the two-spinon continuum of the magnetic subsystem, the dynamical structure factor of the phonons
does not show a Lorentzian character. It is indeed asymmetrical towards low frequencies (Fig. \ref{s2}).
This behavior should be checked in future experiments.
\section{\label{discussion} Discussion}
In this paper we have shown that 
a simple model including 
the elastic frustration of the bilayer structure
is able to produce the incommensurate phase as seen in TiOCl.
On the other hand, from an ab-initio DFT theory,
\cite{zhang} it was recently proposed that a weak {\it ferromagnetic}
inter-chain coupling could be the origin of the incommensuration
in this compound. As no quantitative estimation of the incommensurate wave vector
has been done, we cannot compare the
relative importance of each mechanism. More importantly,
the actual values and the signs of the inter-chain exchanges remain
as a controversial question and the results of the
present work show that the elastic frustration in the interchain coupling
is enough to produce the incommensurate phase.
\newline

In summary, we have developed a model which explains the
uniform-incommensurate transition in TiOCl. It is based on the
competition between the  tendency to dimerize of isolated
antiferromagnetic chains and the frustrated elastic inter-chain
coupling. The model predicts the softening of an incommensurate
mode. Moreover, it is in agreement with the loss of a coherent
spectral weight for the phonons near the zone boundary as observed
by inelastic X-ray scattering.

\begin{acknowledgments}
This work was supported by ANPCyT (PICT 1647), Argentina.
\end{acknowledgments}

\end{document}